\documentclass{article}
\usepackage{spconf,amsmath,graphicx}

\usepackage[english]{babel}
\usepackage[latin1]{inputenc}
\usepackage[T1]{fontenc}
\usepackage{amsmath,graphicx}
\usepackage{amssymb,amsmath,epsfig,url,mathrsfs} \usepackage{algorithm,algorithmic}
\usepackage[usenames, dvipsnames]{color}
\usepackage[framemethod=TikZ]{mdframed}
\usepackage{xcolor}
\usepackage{listings}

\usepackage{ragged2e}
\usepackage{array}

\mdfdefinestyle{MyFrame}{%
    linecolor=black,
    outerlinewidth=.5pt,
    roundcorner=5pt,
    innertopmargin=\baselineskip,
    innerbottommargin=\baselineskip,
    innerrightmargin=5pt,
    innerleftmargin=0pt,
    backgroundcolor=gray!10!white}
\definecolor{codegreen}{rgb}{0,0.6,0}
\definecolor{codegray}{rgb}{0.5,0.5,0.5}
\definecolor{codepurple}{rgb}{0.58,0,0.82}
\definecolor{backcolour}{rgb}{0.95,0.95,0.92}
 
\lstdefinestyle{mystyle}{
    backgroundcolor=\color{backcolour},   
    commentstyle=\color{codegreen},
    keywordstyle=\color{blue},
    numberstyle=\tiny\color{codegray},
    stringstyle=\color{codepurple},
    basicstyle=\footnotesize,
    breakatwhitespace=false,         
    breaklines=true,                 
    captionpos=b,                    
    keepspaces=true,                 
    numbers=none,                    
    numbersep=5pt,                  
    showspaces=false,                
    showstringspaces=false,
    showtabs=false,                  
    tabsize=2
}
 
\usepackage{amsmath}
\usepackage{graphicx}
\usepackage[colorinlistoftodos]{todonotes}
\usepackage[colorlinks=true, allcolors=blue]{hyperref}

\DeclareMathAlphabet{\mathpzc}{OT1}{pzc}{m}{it}

\title{CAN WE USE SPEAKER RECOGNITION TECHNOLOGY TO ATTACK ITSELF? \\ ENHANCING MIMICRY ATTACKS USING AUTOMATIC TARGET SPEAKER SELECTION}

\twoauthors
 {Tomi Kinnunen, Rosa Gonz\'alez Hautam\"aki, Ville Vestman\sthanks{The work was supported by Academy of Finland (project no. 309629).}}
	{School of Computing\\
	University of Eastern Finland, FINLAND\\
	}
 {Md Sahidullah
	}
	{MULTISPEECH\\
	Inria, FRANCE\\
	}

\begin{document}
\ninept
\maketitle
\begin{abstract}
We consider technology-assisted mimicry attacks in the context of automatic speaker verification (ASV). We use ASV itself to select targeted speakers to be attacked by human-based mimicry. We recorded 6 naive mimics for whom we select target celebrities from VoxCeleb1 and VoxCeleb2 corpora (7,365 potential targets) using an i-vector system. The attacker attempts to mimic the selected target, with the utterances subjected to ASV tests using an independently developed x-vector system. Our main finding is negative: even if some of the attacker scores against the target speakers were slightly increased, our mimics did not succeed in spoofing the x-vector system. Interestingly, however, the relative ordering of the selected targets (closest, furthest, median) are consistent between the systems, which suggests some level of transferability between the systems.
\end{abstract}
\begin{keywords}
Speaker verification, mimicry, spoofing
\end{keywords}
\vspace{-0.3cm}
\section{Introduction}
\label{sec:intro}
\vspace{-0.3cm}


It is well known that \emph{representation attacks} ~\cite{Ratha2001,isopad} --- also known as \emph{spoofing attacks} --- cast a shadow over the security of biometric systems. A spoofing attack involves an adversary (attacker) who aims at masquerading oneself as another targeted user to gain illegitimate access. Unprotected \emph{automatic speaker verification} (ASV) systems can easily be spoofed using replay, voice conversion and text-to-speech attacks \cite{Wu2015-spoofing-survey}.
This has sparked research into spoofing \emph{countermeasures} aimed at detecting the attacks from given audio.
Community-driven benchmarks such as ASVspoof \cite{Wu2015-asvspoof} and AVspoof \cite{Ergunay2015-realistic-voice-spoofing} were launched for an organized study of countermeasures. In the context of security, continual arms race between attacks and their defenses is well known \cite{Biggio18-wild}: to develop effective countermeasures, it is necessary to understand the attacks. The speech synthesis community has independently launched \emph{voice conversion challenge}  \cite{Toda2016-VCC-challenge,Lorenzo2018-VCC18} to advance VC methods. Within the past few years, active and dynamic communities both at the `attack' and `defense' sides of ASV, focused on technological attacks, have emerged.


\begin{figure}[!t]
\centering
\includegraphics[width=.98\columnwidth]{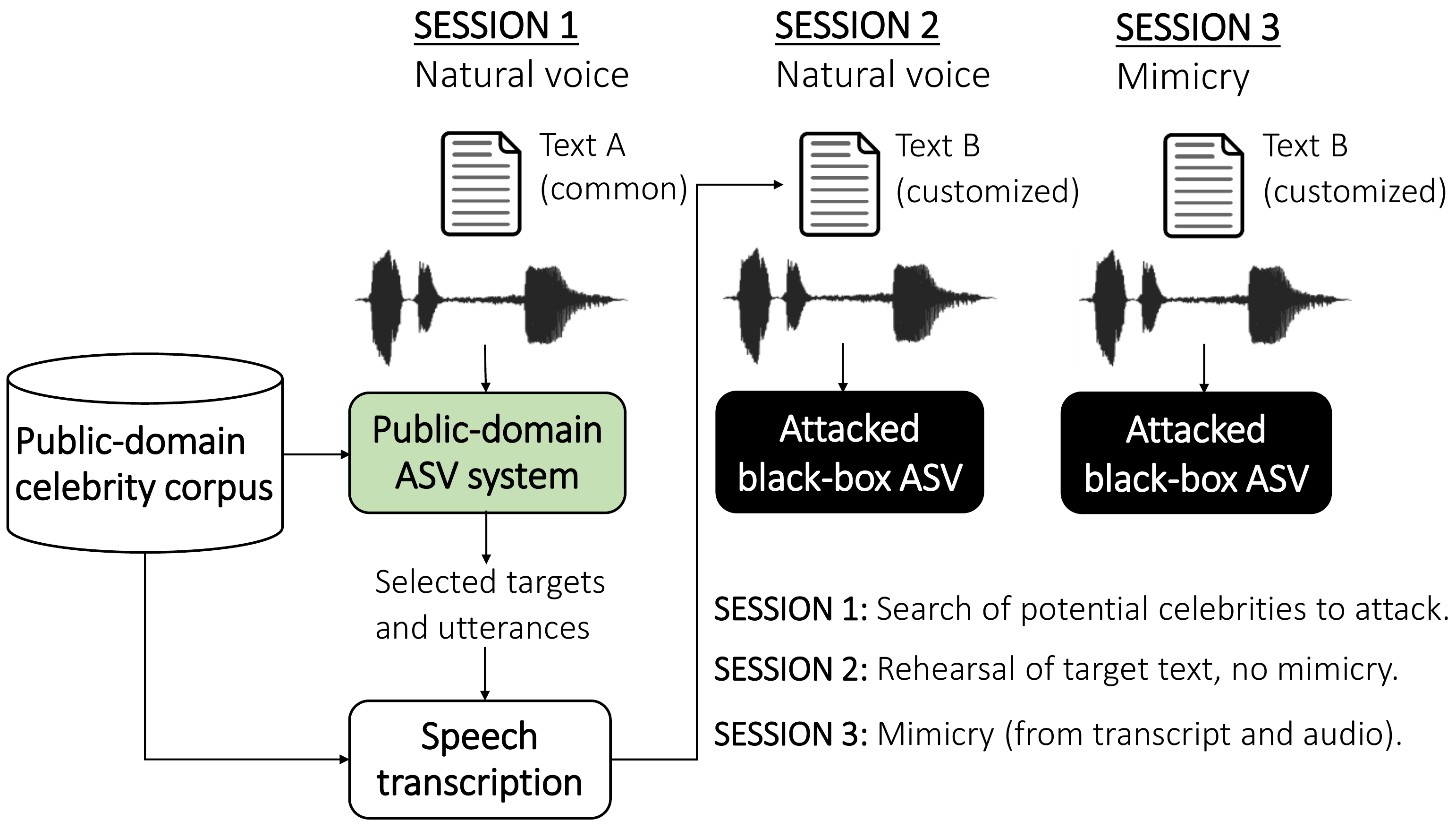}
\label{fig:study-overview}
\vspace{-0.1cm}
\caption{Automatic speaker verification (ASV) assisted mimicry attack: attacker uses a public-domain ASV system to select  target speakers matched with his/her voice from a public celebrity data. The attacker then practices target speaker mimicry, intended to attack another independently developed ASV system.}
\vspace{-0.3cm}
\end{figure}

In this study we focus on a nearly-forgotten ASV attack -- \emph{mimicry} (impersonation). Unlike the technology-induced attacks, mimicry involves \emph{human}-based modification of one's voice production. The question of recognizer vulnerability against mimicry was addressed at least around half a century ago \cite{Luck1969-cepstral,voicespectrograms1971} and has remained a cursory topic within the ASV field \cite{Lau-Vulnerability2004,Lau2005-mimicry,mariethoz2005-professional,eriksson2010disguised,Gonzalez2015-mimicry,Farrus2018}. While ASV vulnerability caused by technical attacks is widely reported, less (reliable) information is available on effectivess of mimicry, due to adoption of small, proprietary datasets. 
The authors are fully aware of the difficulties in collecting mimicry data from professional artists \cite{Gonzalez2015-mimicry}, whose prevalance in the general population is arguably very low. Nonetheless, \emph{if} mimicry attacks could be shown to be a threat to ASV, it would be conceivably challenging to devise countermeasures: natural human speech lacks  processing artifacts that enable detection of technical attacks. Thus, we argue that it is important to keep mimicry also in the list of potential attacks against ASV. Of particular interest in this work are mimicry attacks against \emph{celebrities} whose voice data is available in massive amounts in the public domain. In line with the recent EU's \emph{General Data Protection Regulation} (GDPR), intended to protect the privacy of its citizens, it is important to assess risks associated with multimedia data in the public domain. A recent study \cite{Lorenzo-Trueba2018-can-we-steal} has attempted voice cloning of celebrity voices based on found data using a pre-defined target speaker. The cloned voice samples were, however, detectable as spoofed speech.

We focus on \emph{technology-assisted} mimicry attacks that uses ASV itself to identify potential target speakers to be subjected to mimicry attacks. The idea is to identify targets whose voice is similar to that of the attacker's voice, as this could potentially involve fewer articulatory or voice source modifications. Two related prior studies are \cite{Lau2005-mimicry} and \cite{Panjwani2014-crowd} which involve search of either targets \cite{Lau2005-mimicry} or attackers \cite{Panjwani2014-crowd} from a large set of candidates. The authors of \cite{Lau2005-mimicry} used a Gaussian mixture model (GMM) system to find closest, furthest and median targets from YOHO corpus for a few naive impersonators, leading to substantially increased false acceptance rate. In \cite{Panjwani2014-crowd}, the authors selected impersonators (rather than targets) through a commercial crowd-sourcing platform based on self-judgment and further refinement using ASV. 

Our study can be seen as an attempt to reproduce the findings of \cite{Lau2005-mimicry} using up-to-date ASV technology and a far larger target candidate set ($7,365$ celebrities pooled from VoxCeleb1 \cite{Nagrani2017} and VoxCeleb2 \cite{Chung2018}). A key methodological difference, however, is that unlike \cite{Lau2005-mimicry} that used a \emph{single} GMM recognizer, we include two \emph{different} ASV systems (Fig. \ref{fig:study-overview}). We argue that it may be unrealistic for the attacker to interact many times with the targeted ASV, but he/she may develop an offline  \emph{substitute} ASV that, after optimization, hopefully behaves similar to the attacked one. Our work bears some resemblance to \emph{black box attacks} \cite{Papernot2017-practical-black-box} in adversial machine learning \cite{Biggio18-wild}, though our adversary is not a machine learning algorithm but a human. Further, those methods use either classifier output score or decision to optimize the attacks, while we assume that the attacker receives no feedback from the attacked system in any form. Thus, we expect that such attacks are not strong, but we argue that they are \emph{realistic} given the abundance of multimedia data of celebrity speakers and various public-domain ASV implementations. We seek to answer the question in the title of this work, using a specifically collected `attacker' data and VoxCeleb celebrity targets.


\renewcommand{\arraystretch}{1.0}
\begin{table*}[t!]
\caption{Details of the speaker verification systems used to simulate targeted impersonation attack against automatic speaker verification. The attacker is assumed to not have information about the attacked systems, and hence the attacker's system differs from the attacked systems.}
\label{table:asv-systems}
\footnotesize
\begin{tabular}{p{0.19\linewidth-2\tabcolsep} p{0.27\linewidth-2\tabcolsep} p{0.27\linewidth-2\tabcolsep} p{0.27\linewidth-2\tabcolsep} }
\hline
& \textbf{Attacker's ASV} & \textbf{Attacked ASV 1} & \textbf{Attacked ASV 2}\\
\hline\hline

Sampling rate & 16 kHz & 16 kHz & 8 kHz \\

Acoustic features & 60-dimensional MFCCs (20 static+20-$\Delta$+20-$\Delta\Delta$), RASTA, SAD, CMVN. & 30 MFCC coeffs (no deltas), Sliding CMN normalization & 23 MFCC coeffs (no deltas), Sliding CMN normalization \\

Embedding type & i-vector & x-vector & x-vector \\

Back-end / scoring & PLDA & PLDA & PLDA \\

Development data & WSJ0 and WSJ1, Librispeech & VoxCeleb2, training part of VoxCeleb1 & Switchboard 2 (P. \mbox{1, 2, 3}), Switchboard Cellular, NIST SREs 04 -- 10, Mixer 6 \\

Data augmentation & None & Reverberation, noise, music, babble & Reverberation, noise, music, babble \\

EER (VoxCeleb1) & 12.84 (\%) & 3.11 (\%) & 9.91 (\%)
\\
\hline
\end{tabular}
\end{table*}
\renewcommand{\arraystretch}{1.0}


\vspace{-0.3cm}
\section{ASV-assisted mimicry attacks}
\vspace{-0.2cm}
\subsection{Attack implementation}
\vspace{-0.1cm}

Let $\mathcal{T}=\{T_j\}_{j=1}^J$ denote a set of unique, publicly known \textbf{target speaker} identities and let $\mathcal{A}=\{A_k\}_{k=1}^K$ denote a set of \textbf{attacker} identities. Given a pair of arbitrary utterances (or a pair of collections of many utterances), $(U_i,U_j)$, a \textbf{automatic speaker verification} (ASV) system (speaker detector), $\mathcal{D}(U_i,U_j)$ computes a score $s \in \mathbb{R}$ with an arbitrary scale but higher relative values indicating stronger support that the source of $U_i$ and $U_j$ is the same speaker.

%
We consider two different types of ASV systems. The first one, \textbf{attacker's ASV}  ($\mathcal{D}_\text{pub}$), 
is a public-domain, known ASV implementation, and the latter, \textbf{black-box ASV} ($\mathcal{D}_\text{black}$), is the system an attacker attempts to spoof but whose internal workings, scores and decisions are inaccessible to the attacker. 
The proposed attack proceeds as follows:

\begin{mdframed}[style=MyFrame]\label{asv-assisted-attack}
\center{\textbf{ASV-assisted mimicry attack}}
\begin{small}
    \begin{enumerate}
    \vspace{-0.1cm}
        \item Attacker $A \in \mathcal{A}$ records his/her natural voice sample, $\mathcal{U}_\text{nat}$.
        \item $A$ uses $\mathcal{D}_\text{pub}$ to 
        compute scores $\{s_j\}_{j=1}^J$ between $\mathcal{U}_\text{nat}$ and all the targets in a public database. $A$ picks the \textbf{closest target}, $j^*=\arg\max_{j=1}^J \mathcal{D}_\text{pub}(U_\text{nat},U_j)$, where $U_j$ contains the known recordings of $T_j$. 
        \item $A$ continues to use $\mathcal{D}_\text{pub}$ to pick the best-matching utterances of $T_{j^*}$.
        \item The attacker listens to the selected utterance(s) and attempts to adjust his/her voice towards the target. Once completed practicing, $A$ submits a mimicked test utterance $U_\text{mimic}$ to $\mathcal{D}_\text{black}(U_\text{mimic},U_{j^*})$, the aim of being authenticated as $T_{j^*}$. 
    \vspace{-0.1cm}
    \end{enumerate}
\end{small}
\end{mdframed}

\vspace{-0.1cm}
\subsection{Public-domain (attacker's) ASV system}
\vspace{-0.1cm}
The attacker's ASV uses i-vector front-end and probabilistic discriminant analysis (PLDA) back-end to compute speaker similarity scores. We extract 20 mel-frequency cepstral coefficients (MFCCs) using 20 filters\footnote{\url{http://cs.joensuu.fi/~sahid/codes/AntiSpoofing_Features.zip}}, leading to 60 features per frame after including deltas and double-deltas. The features are processed with RASTA filtering and cepstral mean and variance normalization (CMVN). Non-speech frames are omitted using energy-based speech activity detector (described in Section 5.1 of ~\cite{kinnunen2010overview}).

%

To train the i-vector extractor, we compute sufficient statistics using a universal background model (UBM) of 512 Gaussians. To train the UBM, we choose randomly 10,000 speech utterances from SI-284 subset of the Wall Street Journal corpus (WSJ0 and WSJ1) and 10,000 speech utterances from \texttt{train-clean-360} and \texttt{train-clean-100} subsets of the Librispeech \cite{Panayotov2015-librispeech} corpus. To train the T-matrix with 400 total factors, we randomly choose 20,000 speech utterances from the same subset of WSJ and 20,000 speech utterances from the same subset of Librispeech. Finally, the PLDA is trained on the entire SI-284 subset (from WSJ0 and WSJ1) and entire train-clean subset (from Librispeech) consisting 169,969 speech utterances from a total 1,455 speakers. The 400-dimensional i-vectors are further reduced to 250 dimensions with linear discriminant analysis (LDA) using the same data as in PLDA training, followed by centering and whitening. We use a simplified PLDA with 200-dimensional speaker subspace. We adopt MATLAB-based MSR Identity Toolkit\footnote{\url{https://www.microsoft.com/en-us/download/details.aspx?id=52279}} to train the attacker's ASV.

The above development data and parameter selections are based on preliminary ASV experiments on the AVOID corpus (collected in \cite{GonzalezHautamaki2017-acoustical}) and VoxCeleb (1 and 2). For the AVOID corpus, we obtained EERs of 3.30\% and 4.52\% for text-dependent scenario with two English sentences separately. Further, we obtained EERs of 10.50\% and 16.98\% for text-independent ASV on two subsets of VoxCeleb1 and VoxCeleb2. These custom protocols were created by randomly choosing 60 speakers from VoxCeleb1 (6,163 target and  363,617 non-target trials) and VoxCeleb2 (9,118 target and 537,962 non-target trials).
\vspace{-0.1cm}
\subsection{Attacked ASV systems}
\vspace{-0.1cm}
In our experiments, we regard x-vector systems \cite{snyder2018xvector} based on pre-trained Kaldi \cite{Povey_ASRU2011} recipes as ASV systems to be attacked. To emulate the scenario of attacker's limited knowledge of this system, 
the attacker's ASV is made intentionally different from the attacked ASV systems in terms of feature extractor set-up, embedding type, and development corpora (Table \ref{table:asv-systems}). Both attacked systems are based on Kaldi recipes for VoxCeleb and NIST Speaker Recognition Evaluation 2016, while the attacker's system uses i-vectors.

\vspace{-0.2cm}
\section{Corpus of target speakers: VoxCeleb}
\vspace{-0.2cm}
The attacker's ASV is used as a voice search tool to find the closest speakers from the combination of VoxCeleb1 \cite{Nagrani2017} and Voxceleb2 \cite{Chung2018} to each of the locally recruited subjects (described in Section \ref{sec:local-attackers}). The combined VoxCeleb corpus contains about $1.3$ million speech excerpts extracted from more than 170,000 YouTube videos from $J=7,365$ unique speakers. This totals to about 2,800 hours of audio material that is, for the most part, active speech. Both VoxCeleb corpora were collected using automated pipeline exploiting face verification and active speaker verification technologies \cite{Chung2018}.

VoxCeleb1 contains mostly English speech, while VoxCeleb2 is more diverse in nationalities and languages. The nationality information of the target speakers was of our interest, as the recruited local speakers are Finnish and we wanted to see if Finnish people do better job at imitating Finnish targets than non-Finnish. According to the VoxCeleb1 metadata, there were no Finnish targets in VoxCeleb1. The VoxCeleb2 did not include nationality metadata, so we made a script to automatically obtain nationalities using Google's \emph{Knowledge Graph} API\footnote{\url{https://developers.google.com/knowledge-graph/}}. With this strategy we found 44 Finnish speakers from VoxCeleb2.
\vspace{-0.3cm}
\section{Locally recruited attackers}\label{sec:local-attackers}


\vspace{-0.1cm}
\subsection{Speakers and recording gear}
\vspace{-0.1cm}
We recruited $K=6$ voluntary local speakers (4M + 2F) to serve as `attackers'. All are native Finnish speakers and one of them is a co-author of this study. All the six speakers, selected based on their availability, took part in 2015 to the recordings of \cite{GonzalezHautamaki2017-acoustical}, currently under release with the name AVOID corpus\footnote{\url{http://urn.fi/urn:nbn:fi:lb-2018060621}}. We adopt the same recording setup and part of text prompts from  \cite{GonzalezHautamaki2017-acoustical} but otherwise the two studies are unrelated. 
All the subjects signed an informed consent form to use their speech data for research, and were rewarded with movie and coffee tickets. Two of the male subjects knew the specific goals of our study while the remaining four subjects were not informed that the text and target speakers were tailored for them, nor where do the voices originate from. They were not informed that the study relates to ASV vulnerability, but were merely asked to mimic the targets as accurately as they could.


As illustrated in Fig. \ref{fig:study-overview}, the subjects took part to three recording sessions. The first session, produced in the subject's natural voice, is used for VoxCeleb target speaker selection, while the remaining two sessions serve for vulnerability analysis of the attacked systems. The tasks in the recording sessions differed, while the recording set-up was the same: recordings took place in a silent laboratory room with a portable Zoom H6 Handy Recorder using an omnidirectional headset mic (Glottal Enterprises M80) with 44.1 kHz sampling and 16-bit quantization. Three other channels (two smartphones, electroglottograph) were also collected, but are not used in this study.

\vspace{-0.1cm}
\subsection{The first recording session (data for target search)}
\vspace{-0.1cm}

The first session, used for the targeted VoxCeleb speaker search, consists of four tasks in the speaker's natural voice. The tasks consisted of spontaneous speech and read text (13 sentences) in both Finnish and English. The read texts in Finnish are the same used in \cite{GonzalezHautamaki2017-acoustical} and their corresponding English versions were added for this study. We have approximately 6 minutes of speech (before speech activity detection) per speaker from Session 1. 

\vspace{-0.1cm}
\subsection{Attacked target speaker search and utterance selection}
\vspace{-0.1cm}
\label{sec:selection}
For the purpose of targeted speaker search, we compute a single averaged i-vector for each of the six speakers resulting from 28 individual utterances from Session 1. 
%
%
Similar to \cite{Lau2005-mimicry}, we use the ASV system to pick for each attacker the \textbf{closest}, \textbf{median}, and \textbf{furthest} speakers among the VoxCeleb speakers. The closest one is most relevant for vulnerability analysis while the other two serve for reference purposes. We do this ASV-assisted search separately for \emph{all} the VoxCeleb speakers (unconstrained search from 7,365 speakers) and for the subset of 44 Finnish speakers. We pool all the speech data of the VoxCeleb speakers to compute average i-vector per target. 

In addition to the three ASV-selected targets, we include \textbf{common target} matched with the speaker's gender, in both Finnish and English. The common Finnish targets are P{\"a}ivi R{\"a}s{\"a}nen (F, politician) and Ilkka Kanerva (M, politician), and the common English targets are Hillary R. Clinton (F, politician) and Leonardo DiCaprio (M, actor). 
Even if well-known persons, from the viewpoint of ASV they are \emph{random} targets with no strong presuppositions how similar their voices are to our attackers.

In summary, for each of our 4 male and 2 female subjects, we select 6 customized targets (3 ASV-ranks $\times$ 2 languages) and 2 common gender-matched ones (one Finnish, one English). This gives a theoretical total of $3 \times 2 \times 4 \,\text{male} + 2\,\, \text{common male}$ + $3 \times 2 \times 2\,\, \text{female} + 2\,\, \text{common female} = 40$ target speakers. Not all of the ASV-selected targets are unique, however: one male Finnish celebrity was the closest target for three attackers, one male Finnish celebrity repeated as the median speaker for two male attackers, and one female Finnish celebrity is the furthest speaker for the two female attackers. The final number of unique celebrity targets is 36.  

For each of the 36 target speakers, we selected at minimum 30 seconds of active speech to evaluate the ASV system attacks. This duration of speech was collected from multiple shorter utterances for two reasons: First, the segments in VoxCeleb corpora are typically about 5 to 10 seconds long. Secondly, as we were going to ask the recruited attackers to imitate the target speakers, shorter utterances would be easier to impersonate.

We utilized Attacker's ASV also for the utterance selection. For the closest targets, we selected the highest scoring utterances, while for the furthest targets, the utterances with the lowest scores where selected. For the median speakers, we selected the utterances close to mean. This was further accompanied by manual inspection: if the audio quality (determined subjectively by listening) was not deemed high enough, we discarded it and moved on to the next ones in the ranked list.

\begin{figure*}[t!]

\centerline{\includegraphics[width=\linewidth]{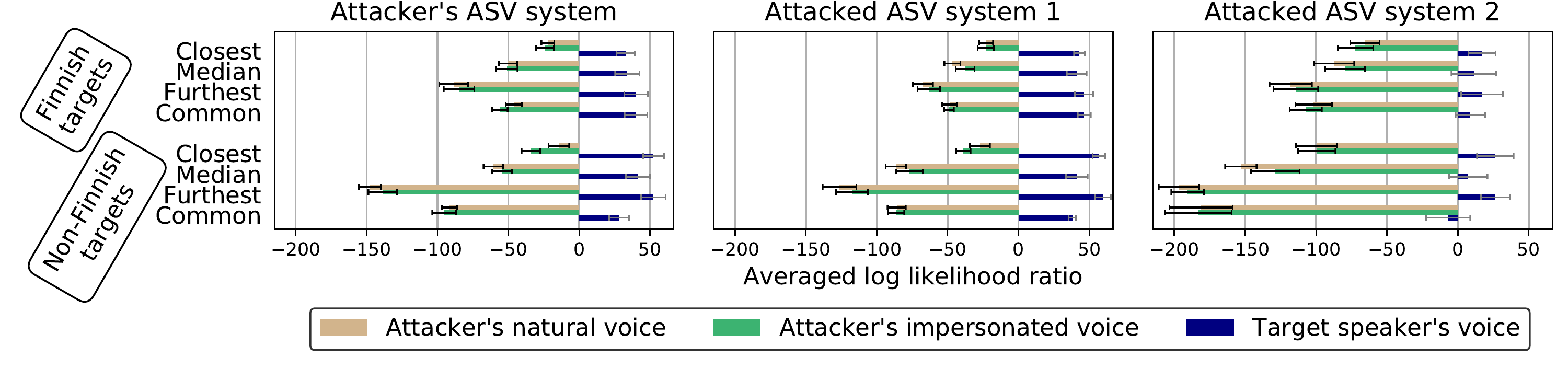}}

\vspace{-6mm}

\caption{Comparison of attackers' ASV scores (log likelihood ratios) to the targets' scores for all the three ASV systems involved in the study. The scores are averaged over all attackers and all speech segments. The error bars represent 95 \% confidence intervals for the means.}
\label{fig:res}
\vspace{-0.3mm}
\end{figure*}

\vspace{-0.1cm}
\subsection{Speech transcription and the mimicry recordings}
\vspace{-0.1cm}
Unlike the first recording session (common to all subjects), the second and third sessions were customized for each subject. This process involves the use of speech transcripts of the selected target utterances. To this end, we used Amazon's Mechanical Turk\footnote{\url{https://www.mturk.com/}} (MTurk), a commercial crowdsourcing service, to transcribe the English language audio. The Finnish transcripts were produced by two native Finnish speakers. The 35 MTurk crowdworkers and the 2 Finnish transcribers were asked to transcribe all the nuances of conversational speech, including repetitions, hesitations, filler words \emph{etc}. Finally, two reviewers audited the quality of all the transcripts. 

In Session 2, which took place 5 to 6 weeks after Session 1, the subject was provided with the  transcripts of the selected target utterance(s) and was asked to read the sentences twice in his or her natural voice. The speaker was not informed whose speech the transcripts corresponded to; even the coauthor taking part to the study was unaware whose transcripts he was reading. The rationale of including this session was to familiarize each attacker with the target speaker sentences. We adopted the general idea to include a session with reference text only and another one with audio from the design used in \cite{mariethoz2005-professional}.

In the last session, which took place 2 to 6 days after Session 2, the subjects were provided with the same transcript as in Session 2, but in addition they had now access to the actual target speaker audio excerpts played through headphones. The transcripts were provided on a printed paper and the audio reference from a tablet computer that the subject was able to interact with; he/she could play the target utterance(s) as many times as needed, and he/she then tried to mimic the voice according to their best skills. Again, the subject was asked to mimic each sentence twice. In the experiments, we use only the second recording of each sentence.

\vspace{-0.2cm}
\section{Results}
\vspace{-0.2cm}

In the following, we evaluate effectiveness of the mimicry attacks. The target speaker models used in the experiments were enrolled using all available segments except those selected for testing as described in Section~\ref{sec:selection}.

Figure \ref{fig:res} displays how the attacker's PLDA scores compare to the  target scores. The general findings are as expected. First, the order of the closest, the median, and the furthest speakers transfers from the attacker's ASV system to the attacked ASV systems implying that the ASV-assisted speaker selection \emph{can} help in ASV attacks. Second, in general, the mimicry attempts were not successful as the attacker's natural and mimicry scores are significantly (and substantially) below the target scores. Additionally, we find no significant difference between the natural and impersonated versions. Finally, as the recruited attackers were Finnish, attackers' scores against the Finnish targets were higher than for non-Finnish targets. 

\begin{figure}[h]

\centerline{\includegraphics[width=\linewidth]{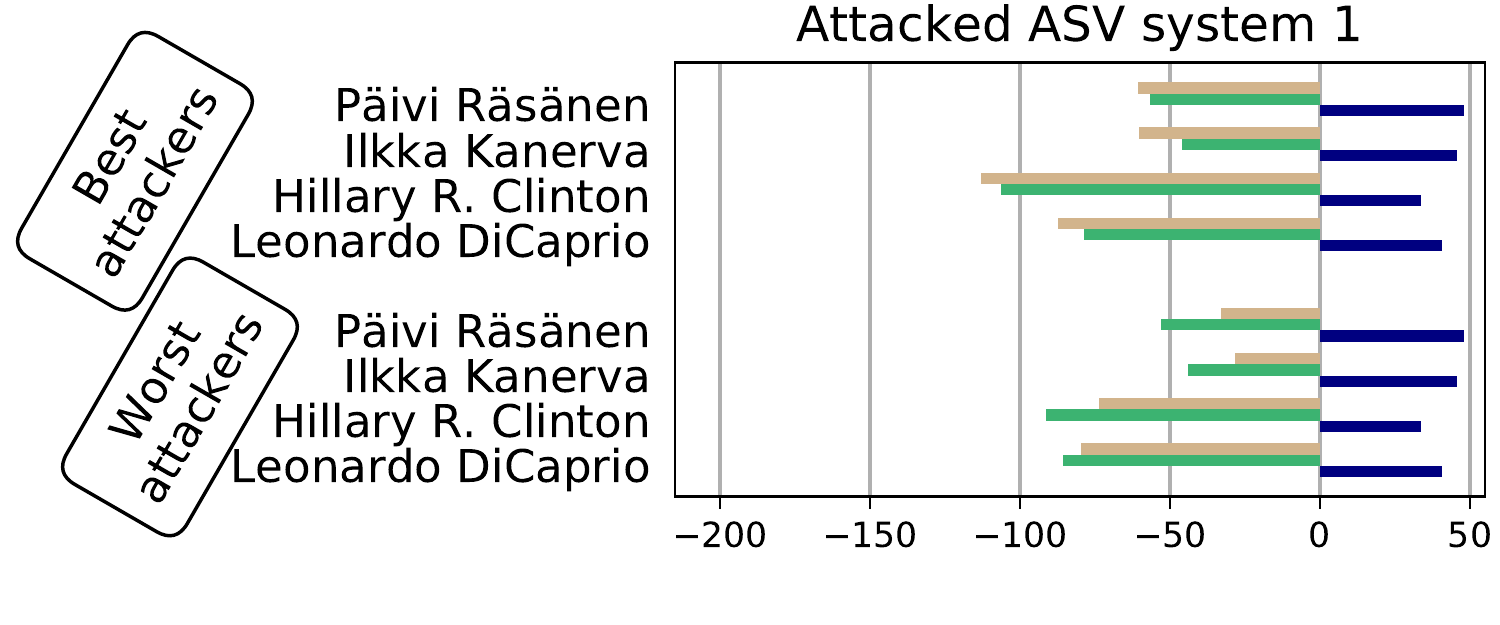}}

\vspace{-6mm}

\caption{Scores against the common celebrity targets for the best and the worst impersonators (Color codes are equal to Fig. \ref{fig:res}).} 

\label{fig:common}
\vspace{-0.3mm}
\end{figure}

To look at the effect of mimicry closer, we analyze the difference of mimicked and natural speech scores (Table \ref{table:mimic}). Interestingly, and contradictory to what we assumed, if the target speaker's voice is already close to the attacker's voice, the impersonation attempts \emph{degrade} the score. The same finding was noted in situations where the target is a well known public figure like the targets in the common category are. We suspect that the effect might be due to people having higher tendency to overact someone they already know well. However, in the case of the targets that are not close to the attackers and, on average, are not so well known (median and furthest categories),  impersonation is helpful. Figure \ref{fig:common} shows a sample of the best and worst attackers for the common targets, with similar findings as above.

\begin{figure}[h]

\centerline{\includegraphics[width=\linewidth]{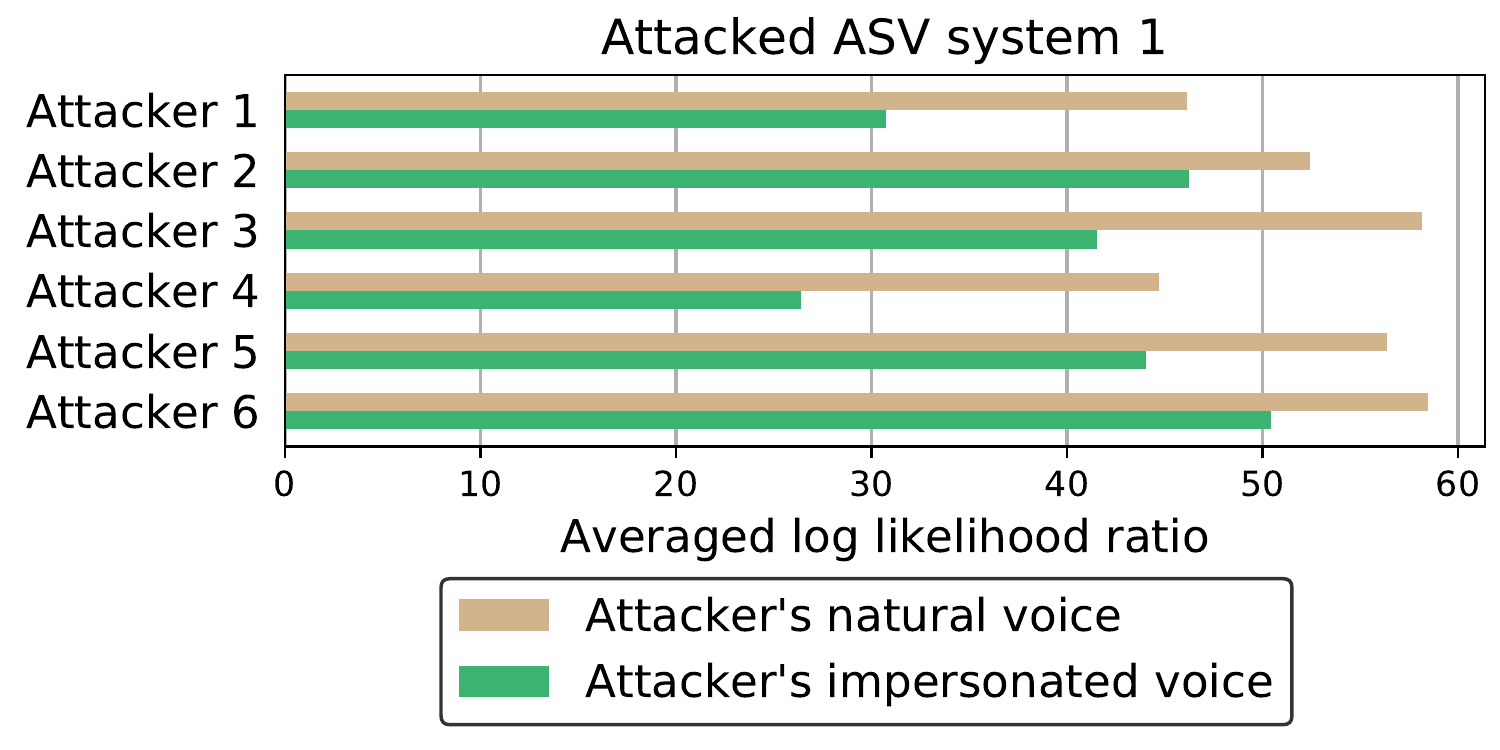}}

\vspace{-2mm}

\caption{Attacker's scores in the case where attacker's test sentences are tested against attackers' own speaker models instead targets' speaker models as in Figure \ref{fig:res}.} 

\label{fig:check}
\vspace{-0.3mm}
\end{figure}

Our attackers are native Finnish speakers recorded with a specific set-up which could be different from the VoxCeleb conditions. This raises a question whether our mimicry attacks were unsuccessful simply due to domain mismatch. To address this concern, we ran an additional experiment, in which we scored attacker's test segments against attacker's own speaker models. If our ASV systems are able to cope with domain mismatch, we expect high detection scores similar to the VoxCeleb target scores). The results shown in Figure \ref{fig:check} confirm this as, similarly to Figure \ref{fig:res}, the averaged log likelihood ratios are close to 50 for the attacked ASV system 1. From Figure \ref{fig:check} we also find that impersonation lowers the scores, showing that the ASV system is not robust against \emph{disguise} (the act of attempting to be not recognized as oneself). This finding was not a surprise to us  \cite{GonzalezHautamaki2017-acoustical}.




\renewcommand{\arraystretch}{1.0}
\begin{table}[t!]
\caption{Score differences between attacks with impersonated voices and attacks with natural voices. Differences are averaged over attackers, target nationalities, and utterances. $\pm$ indicates 95 \% confidence intervals. In the case of the closest target speakers, impersonation attempts are counterproductive.}
\vspace{2mm}
\label{table:mimic}
\small
\begin{tabular}{p{0.20\linewidth-2\tabcolsep}>{\raggedleft\arraybackslash}p{0.20\linewidth-2\tabcolsep}>{\raggedleft\arraybackslash}p{0.21\linewidth-2\tabcolsep}>{\raggedleft\arraybackslash}p{0.19\linewidth-2\tabcolsep}>{\raggedleft\arraybackslash}p{0.20\linewidth-2\tabcolsep}}
\hline
\mbox{ASV system} & Closest & Median & Furthest & Common \\
\hline
Attacker & -9.7 $\pm$ 5.2 & 2.2 $\pm$ 4.3 & 5.9 $\pm$ 7.1 & -7.2 $\pm$ 4.3\\
Attacked1 & -5.2 $\pm$ 3.9 & 9.2 $\pm$ 3.3 & 6.1 $\pm$ 4.3 & -0.5 $\pm$ 3.8\\
Attacked2 & -3.7 $\pm$ 5.5 & 15.0 $\pm$ 7.0 & 4.7 $\pm$ 7.4 & -4.0 $\pm$ 7.7\\

\hline
\end{tabular}
\end{table}
\renewcommand{\arraystretch}{1.0}

\vspace{-0.2cm}
\section{Conclusion}
\vspace{-0.2cm}
Can one use ASV technology to attack itself? 
To answer this, we collected 6 native Finnish mimics and used ASV to locate customized targets from VoxCeleb. Our preliminary analysis reveals that, unlike in \cite{Lau2005-mimicry}, our mimicry attempts were unsuccessful. In fact, the ASV scores even degraded, specifically when the impersonator's natural voice was already close to the target speaker's voice. Further work is required to analyze the reasons, specifically in terms of acoustic modifications implemented by our naive impersonators. The relative ordering of the closest, median and furthest speaker was, however, preserved across the attacker's and attacked ASV systems, with higher relative scores obtained for Finnish targets. Though our assisted attacks did not succeeed to spoof state-of-the-art x-vector technology, selection of imposters from a larger set of speakers (e.g.\, using crowd-sourcing \cite{Panjwani2014-crowd}) may help in spoofing ASV systems. 



\bibliographystyle{IEEEbib}
\bibliography{strings,refs}

\end{document}